**TITLE: Towards Best Practices for Covariate Adjustment in Regulatory Trials: From Fixed to Data-Adaptive Approaches**

**AUTHORS:**

- Laura B. Balzer, Division of Biostatistics, University of California Berkeley, Berkeley, CA, USA
- Lei Nie, Division of Biometrics IV, OB/OTS/CDER/FDA, Silver Spring, Maryland, USA.
- Issa J. Dahabreh, Departments of Epidemiology and Biostatistics, Harvard T.H. Chan School of Public Health and Smith Center for Outcomes Research, Beth Israel Deaconess Hospital, Boston, MA, USA.
- Kajsa Kvist, Novo Nordisk A/S, Denmark.
- Tianyue Zhou, Division of Biostatistics, University of California, Berkeley, Berkeley, CA, USA
- Demissie Alemayehu, Data Science and Analytics, Pfizer Inc., New York, NY, USA.
- Larry Han, Department of Public Health and Health Sciences, Northeastern University, Boston, MA, USA
- Zhiwei Zhang, Biostatistics Innovation Group, Gilead Sciences, Foster City, CA, USA
- Salina P. Waddy, National Center for the Advancement of Translational Science, National Institutes of Health, Bethesda, Maryland
- Andrew Mertens, Division of Biostatistics, University of California Berkeley, Berkeley, CA, USA
- Christian B. Pipper, Biostatistics Methods, Novo Nordisk A/S, Copenhagen, Denmark and Section of Epidemiology, Biostatistics and Biodemography, University of Southern Denmark, Odense, Denmark
- Ken Wiley, Jr., National Center for the Advancement of Translational Science, National Institutes of Health, Bethesda, Maryland, USA
- Margot Yann, Forum for Collaborative Research, School of Public Health, University of California Berkeley, USA
- Gilmer Valdes, OncoBrain Inc and Moffitt Cancer Center, Tampa, Florida, USA
- Xu Shi, Department of Biostatistics, University of Michigan, Ann Arbor, MI, USA
- Mark van der Laan, Division of Biostatistics, University of California Berkeley, Berkeley, CA, USA
- Maya Petersen, Division of Biostatistics, University of California Berkeley, Berkeley, CA, USA
- Kelly Van Lancker, Department of Mathematics, Computer Science and Statistics, Ghent University, Ghent, Belgium and Department of Mathematics and Data Science, Vrije Universiteit Brussel, Brussels, Belgium

Corresponding author: Laura B. Balzer (laura.balzer@berkeley.edu)

**ABSTRACT:**
While randomization justifies the use of unadjusted effect estimators in randomized trials, there is growing interest in covariate adjustment to improve precision. Adjusting for baseline variables that are prognostic of the outcome can reduce estimator variance, resulting in narrower confidence intervals and increased statistical power. Recent guidance by the U.S. Food and Drug Administration supports fixed adjustment for prognostic covariates using parametric regression models. However, this guidance does not address more flexible approaches using data-adaptive or machine learning methods. We offer our perspectives on covariate adjustment to improve analytic precision. We focus on estimating the average effect for the target population in trials with minimal outcome missingness. We provide a non-technical overview of effect estimators that are unadjusted and effect estimators using fixed versus data-adaptive adjustment. We offer practical suggestions for conducting adjusted analyses that are data-adaptive, fully pre-specified, transparently and reproducibly implemented, robust to model misspecification, and guaranteed to improve precision relative to unadjusted analyses – all while preserving statistical validity and the causal effect of interest. We hope that sharing our perspectives will foster broader discussion and eventual acceptance of principled, pre-specified, data-adaptive covariate adjustment in randomized trials.

## INTRODUCTION

Randomized controlled trials are widely considered the preferred approach for evaluating the efficacy of new medical treatments or devices, because randomization helps ensure that persons assigned to different treatment groups are comparable on both measured and unmeasured factors at baseline. In settings with minimal missingness on the outcome, randomization is often sufficient to support unbiased estimation with unadjusted effect estimators. Specifically, we can simply compare the average outcomes between treatment groups. However, statistical efficiency can be meaningfully improved by adjusting for baseline characteristics that are associated with the outcome.[1–4] This approach is known as “covariate adjustment” and makes better use of available data without undermining the validity of the trial. Covariate adjustment can lead to narrower confidence intervals and increased statistical power — helping researchers and regulators draw more precise conclusions from a given trial.

There is a rich body of work on covariate adjustment to improve the precision of trial analyses. This literature starts with regression-style approaches in the early 1930s and extends to recent advances harnessing modern machine learning methods.[5–10] Furthermore, regulatory authorities including the European Medicines Agency (EMA) and the U.S. Food and Drug Administration (FDA) as well as the International Council for Harmonisation (ICH) have issued guidance and recommendations for using covariate adjustment in trial analyses.[11–14] Indeed, the recent FDA guidance supports covariate adjustment as a valuable tool to improve the precision of trial effect estimates when pre-specified and appropriately implemented.[14]

As a group of biostatisticians and clinical trialists from academia, industry, regulatory agencies, and federal research institutions, we view covariate adjustment approaches as valuable tools for improving the precision of trial analyses without compromising validity. However, there are open questions and challenges related to their implementation. Here, we share our perspectives on the use of covariate adjustment to improve precision in regulatory science, overall and particularly when implementing such adjustment using data-adaptive or machine learning methods. Our exposition is intentionally non-technical to reach a diverse group of stakeholders involved in regulatory science. Throughout, we defer technical details to the Appendices and cited references. This work grew out of the Forum on the Integration of Observational and Randomized Data (FIORD) meeting, held in November 2024 in Bethesda, Maryland, USA.

The remaining article is organized as follows. We first provide the scope of our overview: trials with minimal outcome missingness and with a goal of estimating marginal (i.e. population-average) effects. We then review how to obtain marginal effect estimates without adjustment and with fixed adjustment for prognostic covariates. Next, we provide the rationale for data-adaptive approaches to covariate adjustment and discuss some possible implementations. Then, we offer our suggestions for applying data-adaptive adjustment in practice and address common concerns, including precision losses, lack of reproducibility, and poor transparency. We close with a brief discussion.

### *SCOPE OF THIS OVERVIEW*

Throughout, we focus on a frequentist approach to covariate adjustment and the target population framework, in which trial participants are viewed as a random sample from a larger population of interest. This perspective is consistent with recent FDA guidance and facilitates inference that generalizes beyond the study sample.[14] Alternative inferential paradigms — including Bayesian methods, randomization-based inference, or finite-sample methods — may also be appropriate in certain settings and provide complementary approaches (e.g., [15–18]). Altogether, the choice of inferential paradigm should align with the trial design and intended interpretation of uncertainty.

We also focus on clinical trials that are intended to support regulatory decisions and have high measurement coverage for the outcome. In such settings, the motivation for covariate adjustment is to improve precision and statistical power (rather than account for missing data). We also consider an estimand aligned with the intention-to-treat principle, which captures the effect of assignment to the intervention group versus the assignment to the control group, regardless of post-randomization events.[13] We recognize that other settings, including more pragmatic trials, may involve substantial missingness and/or different research objectives, such as estimating effects that account for adherence to the assigned intervention. Here, our goal is to clarify key ideas regarding covariate adjustment in settings where the main inferential goal is improving efficiency rather than reducing bias and where key assumptions are largely justified by the trial design: treatment assignment is randomized and persons with measured outcomes are representative of persons with missing outcomes.

Throughout, we focus on estimation and inference for marginal (i.e., population-average) effects. Let $Y^1$ denote the counterfactual outcome under assignment to the intervention group $(A = 1)$ and $Y^0$ denote the counterfactual outcome under assignment to the control group $(A = 0)$. The quantities $\mathbb{E}(Y^1)$ and $\mathbb{E}(Y^0)$ represent the average outcomes that would be observed if everyone in the population were assigned to the intervention group and to the control group, respectively. Marginal effects are obtained by contrasting these averages on the scale of interest. For example, the marginal effect can be defined on the additive scale as $\mathbb{E}(Y^1) - \mathbb{E}(Y^0)$ or on the relative scale as $\mathbb{E}(Y^1)/\mathbb{E}(Y^0)$.

Marginal effects are distinct from conditional effects, which contrast average outcomes by assignment groups among individuals with the same values of baseline covariates. This distinction concerns the target of inference and should be distinguished from the choice of modeling approach. Marginal and conditional effect measures obtained from nonlinear models are generally not equal because of non-collapsibility.[19] While conditional estimates are often obtained from nonlinear models including terms for the assignment group and baseline covariates, many of these models can also be used to estimate marginal effects. We return to this point below. Although both types of effects can be scientifically meaningful, they answer different questions. Throughout, we focus on marginal effects because they often align most closely with regulatory and clinical decision-making.[11]

## THE UNADJUSTED ANALYSIS

For estimation and inference of marginal effects, suppose that we have conducted a two-group trial with $N$ participants randomized to the intervention group or to the control group. Let $A$ be an indicator of being randomized to the intervention group, and $Y$ be the outcome of interest, measured at the end of the study. For simplicity, we focus on binary, categorical, continuous, and count outcomes. Due to randomization, the average *counterfactual* outcome under assignment level $a$, denoted $\mathbb{E}(Y^a)$, equals the average *observed* outcome under assignment to group $a$, denoted $\mathbb{E}(Y|A = a)$. As a result, marginal (intention-to-treat) effects are identified from the trial data by comparing average outcomes between the randomized groups. Specifically, the marginal effect is identified as $\mathbb{E}(Y|A = 1) - \mathbb{E}(Y|A = 0)$ on the additive scale and as $\mathbb{E}(Y|A = 1)/\mathbb{E}(Y|A = 0)$ on the relative scale.

A natural point estimate of these quantities is the “unadjusted estimator”, which contrasts mean outcomes in the two randomized groups. The group-specific mean outcome is estimated by taking the average outcome observed among participants assigned to that group: $\widehat{\mathbb{E}}(Y|A=a) = \frac{\sum_{i=1}^{N} 1(A_i=a)Y_i}{\sum_{i=1}^{N} 1(A_i=a)}$ where $1(A_i = a)$ is an indicator function, equaling 1 if participant $i$ is in group $a$ and 0 otherwise. The corresponding estimators of the marginal effects are $\widehat{\mathbb{E}}(Y|A=1) - \widehat{\mathbb{E}}(Y|A=0)$ on the additive scale and $\widehat{\mathbb{E}}(Y|A=1)/\widehat{\mathbb{E}}(Y|A=0)$ on the relative scale. Due to random assignment of treatment groups in trials, these estimators are consistent for the marginal effects. Consistent estimators converge to their targets as sample size grows.[20] We refer to **Appendix A** for details on statistical inference, including variance estimation, construction of confidence intervals, and hypothesis testing.

The unadjusted estimator is widely accepted and often the default analysis in randomized trials. We now consider how baseline covariates can be incorporated into the analysis to improve precision, while preserving the target of inference and maintaining statistical validity. We begin with an overview of approaches based on "fixed" adjustment, where the adjustment covariates and functional form are unresponsive to the trial data. Then we discuss data-adaptive approaches to covariate adjustment.

## FIXED ADJUSTMENT

Let $W$ denote baseline covariates that are measured prior to randomization or, at a minimum, not influenced by the trial group. Furthermore, $V$ is a subset of baseline covariates $W$ that are believed to be highly prognostic of the outcome. For example, $V$ might include the outcome measure at baseline (e.g., blood pressure at baseline when the primary endpoint is blood pressure at endline). Alternatively, $V$ may include a "prognostic score" (scalar) derived from historical or external data.[21,22] For a wide variety of outcome types, a more precise estimator of the marginal effect can be obtained by incorporating the prognostic covariates $V$ through a procedure commonly referred to as "G-computation" or "standardization".[1,14,23]

A point estimate is obtained as follows, where we consider a single covariate $V$ for simplicity.

1. Regress the outcome $Y$ on randomized assignment $A$ and the prognostic covariate $V$ according to a pre-specified model for the conditional mean outcome $\mathbb{E}(Y|A,V)$. Common choices include generalized linear models (GLMs) fit by maximum likelihood estimation, such as linear regression for continuous outcomes, logistic regression for binary outcomes, and a log-linear regression for count or rate outcomes. The regression model must include an intercept and a main term for $A$. For example, we could specify a main-terms GLM as $g\{\mathbb{E}(Y|A,V)\} = \beta_0 + \beta_1 A + \beta_2 V$ where g{·} denotes the canonical link function (e.g., the identity, logit, or log link). Equivalently, we can express this regression model as $\mathbb{E}(Y|A,V) = g^{-1}(\beta_0 + \beta_1 A + \beta_2 V)$.

2. Using the regression fit, predict the outcome for all $N$ participants under assignment to the intervention group and under assignment to the control group: $\widehat{\mathbb{E}}(Y|A=1,V_i) = g^{-1}(\hat{\beta}_0 + \hat{\beta}_1 + \hat{\beta}_2 V_i)$ and $\widehat{\mathbb{E}}(Y|A=0,V_i) = g^{-1}(\hat{\beta}_0 + \hat{\beta}_2 V_i)$ for $i = 1, \ldots, N$.
3. Average these predictions across all $N$ participants to obtain marginal mean estimates under each assignment strategy: $\frac{1}{N}\sum_{i=1}^{N} \widehat{\mathbb{E}}(Y|A=1,V_i)$ versus $\frac{1}{N}\sum_{i=1}^{N} \widehat{\mathbb{E}}(Y|A=0,V_i)$. Then contrast on the scale of interest.

We refer to **Appendix A** for an overview of obtaining statistical inference.

The regression model in Step 1 is viewed as a "working" model, meaning that it does not need to represent the true data-generating process.[2] Indeed, due to randomization of the treatment group, approaches using covariate adjustment are quite robust to the model misspecification. As an illustration, suppose the true outcome regression contains an interaction between the assignment *A* and the prognostic covariate $V$: $\mathbb{E}(Y|A,V) = g^{-1}(\beta_0 + \beta_1 A + \beta_2 V + \beta_3 AV)$. If we use a main-terms model omitting the interaction, the resulting G-computation estimator remains consistent for the marginal effect. More generally, consistency follows whenever the prediction unbiasedness condition holds; that is, when the average of the predicted outcomes equals the average of the observed outcomes within each randomized group: $\frac{1}{N}\sum_{i=1}^{N} \widehat{\mathbb{E}}(Y|A=a,V_i) = \widehat{\mathbb{E}}(Y|A=a)$ for $a = \{0,1\}$.[2,24–26] This condition is commonly achieved by specifying a GLM with the canonical link, an intercept, and a main term for the randomized assignment and then fitting the regression with maximum likelihood or quasi-likelihood estimation.

Importantly, as the outcome regression model more closely approximates the true conditional expectation, the resulting G-computation estimator becomes more precise. Under correct specification of the outcome regression model, the estimator is "locally efficient", yielding the tightest confidence intervals and most statistical power.[1,2] We note that if the prognostic covariates $V$ contain missing values, simple approaches such as mean or mode imputation can be used, because baseline covariates are included solely to improve precision, rather than to remove bias.[27,28]

The G-computation approach to covariate adjustment focuses on improving precision by estimating the conditional mean outcome $\mathbb{E}(Y|A,V)$. Alternative approaches such as inverse weighting exploit information in the assignment mechanism, which is often summarized through the "propensity score": the conditional probability of being randomized to the intervention group given the covariates, denoted $\mathbb{P}(A=1|V)$. In observational studies, propensity score methods are commonly used to account for confounding. In randomized trials, there is no confounding and the assignment probability is known; so, estimation of the propensity score is not required. Nevertheless, in finite samples, estimating the known propensity score $\widehat{\mathbb{P}}(A=1|V)$ in trials can improve precision by accounting for chance imbalances on predictive covariates by group.[29–31]

Alternative estimators, such as augmented inverse probability weighting (AIPW) and targeted minimum loss-based estimation (TMLE), combine estimates of the outcome regression and the propensity score and can provide additional gains in efficiency relative to approaches relying on only estimating the outcome regression or the propensity score.[29,32–35] For trials, Rosenblum and van der Laan provide a comprehensive overview of these approaches, treating the G-computation estimator as a special case and discussing their relationship to more traditional approaches, such as the analysis of covariance (ANCOVA).[2] Additional overviews can be found in Tsiatis et al. and Van Lancker et al.[1,4] Notably, AIPW and TMLE are often referred to as "doubly robust" estimators, because they will yield a consistent point estimate if either the outcome regression or propensity score model is correctly specified. In randomized trials where the propensity score is known, these approaches are fully model-robust. Therefore, we can again use "working" models for the outcome regression and propensity score to improve precision.

Thus far, we have considered "fixed" approaches to covariate adjustment where the choices of the covariate set $V$ and the functional form of the outcome regression model and/or propensity score model do not incorporate or adapt to the trial data. Approaches based on fixed adjustment may improve precision relative to the unadjusted estimator, but such gains are generally not guaranteed in small or large samples. Indeed, including covariates that are weakly prognostic or unrelated to the outcome can reduce the precision of trial analyses in practice. This precision loss arises because including non-prognostic covariates can increase variability without meaningfully reducing residual outcome variation. In an ideal setting, with perfect knowledge, such covariates would never be included in the adjustment set. In practice, however, the prognostic value of the candidate covariates is typically not known with certainty. Indeed, recent examples have shown that forced adjustment can lead to precision losses relative to the unadjusted effect estimator in finite samples.[8]

More generally, decisions about which covariates to include and how best to adjust for them (i.e., choice of functional form, including use of nonlinear terms or interactions) can have major effects on precision and power. These decisions should be pre-specified; yet, the optimal choices that maximize precision and power are generally unknown before the trial data are collected. This tension motivates the use of data-adaptive approaches, including machine learning methods, that allow the data to inform covariate selection and regression model specification, while preserving valid inference for the same marginal effect.

## DATA-ADAPTIVE ADJUSTMENT

Data-adaptive approaches to covariate adjustment are designed to address these limitations. Instead of pre-specifying a specific regression form, these approaches use the trial data to guide the selection of covariates and the choice of functional form through a pre-specified algorithm. In this way, the adjustment strategy is completely defined before unblinding, while still allowing the selected regression model to depend on the data. Concretely, in the above G-computation procedure, we could replace the fixed regression in Step 1 with stepwise regression to facilitate the selection of covariates with the strongest predictive value and, thereby, improve efficiency over the unadjusted estimator. To further improve efficiency, we could consider more flexible methods, such as multivariate adaptive regression splines (MARS), that can more closely approximate the true conditional mean outcome $\mathbb{E}(Y|A,V)$. To preserve the validity of the estimator for the marginal effect of interest, we still want to ensure the final outcome regression includes an intercept and a main term for the randomized group. In AIPW and TMLE, data-adaptive estimation of the known propensity score can lead to further efficiency gains compared to approaches relying only on the outcome regression.

A foundational principle enabling data-adaptive adjustment is that the treatment groups are randomly assigned. This allows the use of machine learning for flexible estimation to facilitate gains in efficiency, while still providing consistent and asymptotically normal estimates under standard regularity conditions.[20,34] In essence, these conditions require the estimators of the outcome regression and the propensity score do not overfit; overfitting occurs when the estimator is too responsive to noise, versus underlying relationships. Cross-fitting is recommended to protect against overfitting when using data-adaptive approaches, such as penalized regressions, tree-based methods, or ensemble approaches. Cross-fitting refers to a sample-splitting procedure where nuisance functions (i.e., the outcome regression and propensity score models) are trained on part of the data and evaluated on a different part of the data.[36,37]

Importantly, using data-adaptive methods for covariate adjustment does not inflate Type I error, provided the above regularity conditions are met. The variance of these estimators is typically derived via influence curve-based approaches (i.e., sandwich estimators), as outlined in **Appendix A**.[20,30,34,38,39] This approach accounts for the uncertainty introduced by the data-adaptive process by using the empirical variance of the estimated influence curve, rather than a fixed-model likelihood.

While such approaches to data-adaptive adjustment are fully pre-specified, they are still not guaranteed to yield an estimator with lower variance than the unadjusted estimator. Therefore, we suggest using fully pre-specified, data-adaptive procedures with the following three desirable properties. First, they use data collected in the trial to select the adjustment covariate(s) $V$ from the larger set of baseline covariates $W$ in a way that ensures that precision is improved or, at least, not lost as compared to the unadjusted approach for the same estimand. Second, they incorporate flexible modeling of the outcome regression to get closer to the true conditional mean outcome $\mathbb{E}(Y|A,V)$ for maximum gains in efficiency. Third, they yield valid inference under no or minimal additional assumptions.

Four procedures meeting these criteria were developed by members of this working group. Building off the principle of Empirical Efficiency Maximization,[40] Balzer and colleagues introduced TMLE with Adaptive Pre-specification to use sample-splitting to select the adjustment approach that minimized the cross-validated variance estimate.[41] Combining the principles of Empirical Efficiency Maximization and Super Learner,[40,42] Zhang and Ma also proposed a procedure to data-adaptively select the combination of predictions that minimized cross-validated variance estimate.[6] Liu and colleagues introduced COADVISE (covariate adjustment with variable selection), which fits the outcome regression separately for each trial arm and conducts estimation and inference with AIPW.[9] De Bartolomeis and colleagues proposed Hybrid-AIPW (H-AIPW) to improve precision by combining outcome regression estimators fit on data internal to the trial with estimators fit on data external to the trial.[10]

While numerous methodological papers have demonstrated the promise of data-adaptive approaches to covariate adjustment to improve analytic precision, their application in the primary analysis of trials remains limited. Balzer and colleagues summarized the use of TMLE with Adaptive Pre-specification in the primary analysis of 8 recently published trials.[43] These were Phase 4+ trials and were designed to evaluate the implementation and real-world effectiveness of interventions with established efficacy. In these studies, the precision gains from adaptive adjustment instead of the unadjusted estimator ranged from substantial (64% more precise) to none, but with no instances as harm.[43] To the best of our knowledge, there has been limited, if any, use of data-adaptive covariate adjustment in the primary analysis of trials intended to support regulatory decision-making. Instead, data-adaptive approaches to covariate adjustment have more commonly been pre-specified as secondary or sensitivity analyses. In the following, we review several persistent concerns regarding covariate adjustment in randomized trials, overall and particularly when using data-adaptive methods.

## CONSIDERATIONS FOR COVARIATE ADJUSTMENT

Here, we provide practical suggestions on best practices for implementing covariate adjustment in trials intended to support regulatory decision-making. We also highlight areas where further discussion, methodological development, and regulatory guidance may be needed.

### *Stand by Your Estimand*

All trials need a clear definition of the treatment effect (i.e., the causal estimand) and alignment between that effect and the analysis undertaken. This principle is emphasized in the ICH E9(R1) Addendum.[13] For the marginal effect considered throughout, a long-standing concern is that covariate adjustment may change the target of inference from a marginal effect to a conditional effect. This concern is particularly prominent when effects are estimated directly from coefficients in outcome regression models for binary or time-to-event endpoints.[19,44] However, all approaches discussed above incorporate standardization to generate marginal effect estimates, regardless of the outcome-type and even when the outcome regression is non-linear. This is true for both fixed and data-adaptive approaches to covariate adjustment.

### *Avoid Bad Bets*

Recent FDA guidance warned of potential precision losses when adjusting for covariates that are not, in-fact, prognostic of the outcome.[14] This is indeed a legitimate concern in practice. To avoid betting on an adjustment approach that may actually harm precision in finite samples, data-adaptive approaches built on the principle of Empirical Efficiency Maximization offer an attractive alternative.[40] As long as the unadjusted estimator is included as a candidate, the approaches using Adaptive Pre-specification and Super Learner are guaranteed for each trial analysis to be at least as precise as the unadjusted approach for the chosen effect and variance criterion; specifically, they default to the unadjusted approach if none of the candidates using covariate adjustment reduce the cross-validated variance estimate.[6,8,41] Alternatively, under stated conditions and with a calibration step for nonlinear outcome regression models, the COADVISE approach guarantees that the AIPW-based estimator is at least as efficient as the unadjusted estimator asymptotically.[9] Similarly, under mild regularity conditions, H-AIPW guarantees asymptotic efficiency at least as great as that of the estimator using only data internal to the trial (including possibly the unadjusted estimator), even under model misspecification.[10]

### *Pre-specification or Bust!*

The approach for evaluating treatment efficacy must be pre-specified. This principle is emphasized by both the EMA and FDA.[12,14] The trial's Protocol and/or Statistical Analysis Plan must detail the exact algorithm for point estimation and inference. Use of data-adaptive procedures are completely compatible with this recommendation. It is essential to specify the estimand, the candidate approaches to covariate adjustment, as well as the exact procedures for selection among those candidates and for obtaining a final point and variance estimate. This ensures that the analysis is reproducible and not influenced by knowledge of the treatment effect.

A related concern with data-adaptive adjustment is that researchers will "snoop": use the unblinded data to select the analytic approach that maximizes the effect estimate. This danger is avoided when the analytic approach for effect estimation – adjusted or not – is fully pre-specified prior to unblinding. We also recommend that corresponding computing code be "locked" prior to unblinding and prior to examining any post-baseline data by group. We note that this recommendation does *not* imply that covariate or functional form selection must be performed in a way that is blinded to treatment assignment. Blinded selection is unnecessarily restrictive and may forgo potential efficiency gains. As previously discussed, data-adaptive estimation of the outcome regression and known propensity score with the full, unblinded data does not compromise validity of effect estimation or inference. Thus, when properly pre-specified and implemented, unblinded data-adaptive adjustment can safely improve precision rather than degrade it.

***Ready; Set; Reproduce!***

A related concern with covariate adjustment, particularly when using machine learning, is the potential loss of reproducibility. Strict adherence to pre-specification practices can help mitigate this issue, especially when the tuning parameters of all algorithms are explicitly stated. Nonetheless, data-adaptive approaches involving stochastic elements can yield different results when the analysis is rerun. At minimum, we recommend pre-specifying and setting the random seed within the statistical software (e.g., *set.seed()* in *R*), recording the exact versions of all software dependencies (e.g., *renv()* in *R*), using code containerization (e.g., Docker) to save operating system and system libraries, and archiving all computing code. We also recommend making all analytic code publicly available on GitHub or similar platforms. As covariate adjustment becomes more widely adopted, standardized and well-validated software will further support consistent and reproducible analyses. For example, dedicated packages such as *RobinCar2, HAIPW,* and *Coadvise* implement covariate-adjusted estimators within transparent and reproducible workflows – helping investigators apply these methods in a manner consistent with pre-specification requirements.[10,45,46]

***Don't Double-Dip; Cross-fit***

Another persistent concern is "double-dipping", which arises when the same data are used to select the adjustment approach and to evaluate the treatment effect. The core concern is that such an approach leads to variance underestimation and inflated Type-I error rates. Under reasonable conditions, AIPW and TMLE remain consistent and asymptotically normal even when using machine learning, and the standard approach to variance estimation accounts for the uncertainty added due to data-adaptive learning (**Appendices A-B**). In practice, however, it is possible to overfit, especially when attempting to adjust for a large number of covariates and/or when using very adaptive learners. A simple solution is to limit the candidates to working GLMs adjusting for one covariate.[21,22,41] However, this approach might be too conservative in that including more covariates and using more flexible modeling approaches may lead to meaningful efficiency gains.[8,43] On the other hand, use of more flexible methods may require inference based on cross-fitting to separate estimation of nuisance functions (i.e., the outcome regression and the propensity score) from estimation of the treatment effect and to stabilize finite-sample performance.[36,37] As described next, simulations can help inform these decisions.

### *Simulations Can Be Stimulating*

Conducting a simulation study can help to inform trial design (i.e., which baseline covariates $W$ to collect), pre-specify the Statistical Analysis Plan, confirm expected performance of the primary analytic approach prior to unblinding, and convince stakeholders of the benefit of modern analytic approaches. These simulations should be as realistic as possible in that they reflect the key characteristics of the trial, such as the sample size, outcome rarity, as well as the expected number and strength of the covariates.[47] While software or implementation limitations may make it challenging for some research groups to conduct in-depth simulations, we have found the approach invaluable. Furthermore, recent advances in generative AI may facilitate more widespread use of realistic simulations to inform trial design and analysis.[48]

### *Communication Is Key*

We acknowledge and appreciate that buy-in from all stakeholders is core to implementing novel methods in practice. Clear articulation of the estimand, the algorithm, and its expected performance is as important as the choice of the method itself. Furthermore, when machine learning algorithms are used, the resulting analytic procedure can become opaque to readers, reviewers, and regulators, especially if the modeling process is not clearly documented or reported. We recommend that in addition to publishing the Statistical Analysis Plan and analytic code, the main results paper should report the final approach to covariate adjustment. This includes the specific adjustment sets, the functional form of the outcome regression and propensity score model as well as the procedure by which they were selected. If ensemble methods, like Super Learner,[42] are used, the weight given to each algorithm should be provided. We also recommend reporting the results from the unadjusted analysis and the efficiency gains from adjustment. Such disclosures enhance interpretability and facilitate critical appraisal of the analytic choices made.

## CONCLUSION

Covariate adjustment has long been recognized as a principled way to improve precision in randomized trials without compromising the protection afforded by randomization. For over a decade, regulatory agencies such as the EMA have endorsed adjustment for a small set of pre-specified prognostic variables, and recent FDA guidance further clarifies that covariate-adjusted analyses are appropriate — and often desirable — when procedures are defined in advance.[12,14] These statements reflect an emerging consensus that efficiency gains should be pursued whenever they can be achieved without changing the target estimand or compromising the statistical validity. Regulatory guidance emphasizes pre-specification of the analysis strategy, selection of clinically and prognostically relevant baseline covariates, and use of methods that preserve Type 1 error control.

The recent FDA guidance describes the G-computation approach using fixed GLMs, but does not cover doubly robust approaches or the use of machine learning. Indeed, the landscape of available methods has evolved. Restricting adjustment to a fixed and limited set of covariates may leave meaningful efficiency gains unrealized, particularly in large trials with rich baseline or historical data. Approaches like AIPW and TMLE provide a principled way to data-adaptively leverage this information, while maintaining valid inference. In randomized trials, these approaches can be viewed as model-assisted methods using flexible prediction to improve precision, while retaining the design-based protection of randomization. When fully pre-specified, implemented appropriately, and transparently reported, these methods preserve the core requirements of biopharmaceutical research: robustness, integrity, interpretability, reproducibility, and control of Type I error.

In conclusion, we suggest that sponsors, investigators, and regulators view data-adaptive adjustment not as an experimental add-on, but as a natural extension of established practices for covariate adjustment. With thoughtful design, transparent reporting, and continued dialogue across academia, industry, and regulatory agencies, these methods can responsibly enhance the precision and reliability of trial evidence. We also emphasize that our discussion has focused specifically on covariate adjustment for primary efficacy analyses and does not address several related complexities, including censored outcomes and intercurrent events, nor the integration of these methods within adaptive or sequential trial designs. We also focused on covariate adjustment using only the trial data and did not consider approaches integrating external data into the effect estimates (e.g., [49,50]). These areas warrant separate methodological consideration and remain important directions for future work and discussion. Continued methodological research and applied experience will further clarify best practices, but the path forward is clear: principled, pre-specified, data-adaptive adjustment has an important role to play in the next generation of confirmatory clinical trials.

## Appendix A – On Statistical Inference

In the main text, we outlined how to obtain a point estimate with either an unadjusted approach (i.e., by comparing the average outcome in each randomized group) or the G-computation approach. We also discussed doubly robust estimators, such as AIPW and TMLE, which combine estimates of the outcome regression and known propensity score. Due to randomization of the treatment group, all estimators are generally consistent for marginal effects. Here, we outline how to obtain statistical inference.

Suppose, for example, we are interested in the effect on the additive scale and using an influence curve-based (i.e., sandwich) inference.[20,30,34,38,39] For the unadjusted estimator, the estimated influence curve for participant $i$ is

$$\widehat{IC}_i = \frac{1(A_i=1)}{\widehat{\mathbb{P}}(A=1)}\left[Y_i - \widehat{\mathbb{E}}(Y|A=1)\right] - \frac{1(A_i=0)}{\widehat{\mathbb{P}}(A=0)}\left[Y_i - \widehat{\mathbb{E}}(Y|A=0)\right].$$

For G-computation incorporating adjustment for covariate(s) $V$ through the outcome regression, the estimated influence curve for participant $i$ is

$$\widehat{IC}_i = \frac{1(A_i=1)}{\widehat{\mathbb{P}}(A=1)}\left[Y_i - \widehat{\mathbb{E}}(Y|A=1,V_i)\right] - \frac{1(A_i=0)}{\widehat{\mathbb{P}}(A=0)}\left[Y_i - \widehat{\mathbb{E}}(Y|A=0,V_i)\right] + \widehat{\mathbb{E}}(Y|A=1,V_i) - \widehat{\mathbb{E}}(Y|A=0,V_i) - \hat{\varphi}$$

where $\hat{\varphi}$ denotes the point estimate. For doubly robust estimators, such as AIPW or TMLE, incorporating additional adjustment in the propensity score, the estimated influence curve for participant $i$ as

$$\widehat{IC}_i = \frac{1(A_i=1)}{\widehat{\mathbb{P}}(A=1|V_i)}\left[Y_i - \widehat{\mathbb{E}}(Y|A=1,V_i)\right] - \frac{1(A_i=0)}{\widehat{\mathbb{P}}(A=0|V_i)}\left[Y_i - \widehat{\mathbb{E}}(Y|A=0,V_i)\right] + \widehat{\mathbb{E}}(Y|A=1,V_i) - \widehat{\mathbb{E}}(Y|A=0,V_i) - \hat{\varphi}.$$

If we were interested in effects on other scales (e.g., relative effects), the influence curve would have a different form, but the core components would be the same.

For the unadjusted estimator or approaches using fixed adjustment, a variance estimate is obtained from the empirical variance of the estimated influence curve scaled by total sample size $N$:

$$\hat{\sigma}^2 = \frac{\widehat{Var}(\widehat{IC})}{N} = \frac{1}{N^2}\sum_{i=1}^{N}\widehat{IC}_i^2$$

where the last equality holds because the estimated influence curve is centered (i.e., its empirical mean is approximately 0). When using data-adaptive methods, this approach to variance estimation can be used when the standard regularity conditions are met, as discussed in the main text. However, obtaining a cross-fitted inference is often recommended when using data-adaptive procedures. Briefly, the nuisance functions (e.g., the outcome regression and propensity score) are estimated on training samples that exclude the observation being evaluated. The influence curve is then calculated using these cross-fitted estimates, and a variance estimate is obtained from the empirical variance of the cross-fitted influence-curve values divided by sample size $N$.

We then use the point estimate $\hat{\varphi}$ and the variance estimate $\hat{\sigma}^2$ to form Wald-type 95% confidence intervals and to carry out hypothesis tests in the usual way. If the outcome regression is misspecified, it has often been stated that the resulting inference will typically be conservative.[34] However, as discussed in **Appendix B**, recent research suggests that inference for several common estimators is asymptotically exact even when machine learning is used, provided the regularity conditions are satisfied.[7,51]

For many estimators, nonparametric bootstrap methods can also be used to obtain confidence intervals and p-values. Both influence curve-based inference and the nonparametric bootstrap are justified by asymptotic theory. The bootstrap is often convenient to implement and can approximate the sampling distribution of complex estimators without explicit derivation of an influence curve. However, for common estimators, such as the difference in mean outcomes, the influence curve yields a simple closed-form variance estimator (above), which is algebraically equivalent to the usual two-sample variance estimator. We also note that the bias-corrected and accelerated bootstrap sometimes provides improved finite-sample performance as compared to simple Wald intervals.[3]

## Appendix B – Asymptotically exact inference

Recent research demonstrates that influence curve-based variance estimation can provide asymptotically exact inference even when the outcome regression model is misspecified.[7,51] For time-to-event and ordinal outcomes, TMLEs that incorporate covariate adjustment in the outcome model via machine learning methods, such as LASSO, Random Forests, and XGBoost, can achieve asymptotically exact Type 1 error control and valid confidence intervals under very general conditions. In other words, they provide asymptotically exact inference as long as the machine learning models converge to some limit, not necessarily the true conditional mean outcome. Extensions of these results show that combining cross-fitting with bias-robust modeling techniques further ensures asymptotically exact inference across a broader range of endpoints and estimation strategies, effectively bridging regulatory requirements and modern efficiency-oriented procedures. Other AIPW-type estimators that incorporate covariates in the outcome regression model via machine learning methods need a slightly adapted influence curve.[51]

## Acknowledgments:

We would like to thank the sponsors and attendees of the FIORD workshop, including the Forum for Collaborative Research and the Center for Targeted Machine Learning and Causal Inference (both at the School of Public Health at the University of California, Berkeley), and the Joint Initiative for Causal Inference.

## Funding statement:

This research was funded by a gift from the Novo Nordisk corporation to the University of California, Berkeley, to support the Joint Initiative for Causal Inference.

## Disclosures:

The contents are those of the authors and do not necessarily represent the official views of, nor an endorsement by, the FDA/HHS, the US Government, or the authors' affiliations. Indeed, this paper represents the views of the authors and should not be interpreted as FDA policy or guidance beyond what is explicitly stated in the published FDA 2023 guidance document.

The authors report generative AI was not used in their research or preparation of this manuscript).

## Competing Interests:

DA works for and owns stocks in a pharmaceutical company.

CBP is an employee of Novo Nordisk and holds shares in Novo Nordisk.

ZZ is employed by and holds shares in Gilead Sciences, Inc.